\documentclass{jpsj-suppl}
\usepackage{txfonts} %Please comment out this line unless the txfonts package is availabe in your LaTeX system.

\title{ The Dubna-Mainz-Taipei Dynamical Model for $\pi N$ Scattering and $\pi$ Electromagnetic Production}

\author{Shin Nan \textsc{YANG}$^{1}$}
\inst{$^{1}$Department of Physics and Center for Theoretical Sciences, Taipei 10716, Taiwan}
\email{snyang@phys.ntu.edu.tw}

\recdate{October 5, 2015}

\abst{Some of  the featured results of the Dubna-Mainz-Taipei (DMT)  dynamical model  for $\pi N$ scattering and $\pi^0$ electromagnetic  production are summarized.
These include results for threshold $\pi^0$ production, deformation of $\Delta(1232)$, and the extracted   properties
of higher resonances below 2 GeV. The excellent agreement of DMT model's predictions with threshold $\pi^0$ production data, including the recent precision measurements
from MAMI establishes results of  DMT model as a benchmark for experimentalists and theorists in dealing with threshold pion production.}

\kword{$\pi N$ scattering, electromagnetic production of pions, meson-exchange model
}

\begin{document}
\maketitle

\section{Introduction}

Dynamical approach to meson electromagnetic production has been widely employed   to analyze and interpret experimental
data since it was proposed in \cite{Tanabe85,Yang85}. There are now many dynamical models constructed. Currently, the most sophisticated dynamical model
ever constructed is the Argonne-Osaka dynamical coupled-channel model \cite{Kamano13}, as presented by Kamano in this conference \cite{Kamano15}. It
includes eight channels and 487 parameters.

In this talk, I will summarize the results obtained with Dubna-Mainz-Taipei (DMT) meson-exchange dynamical model for $\pi N$ scattering and electromagnetic (EM)
production of pion. It was prompted by the following  motivations. First, to construct a meson-exchange model for $\pi N$ scattering and EM production of pion in order to achieve a unified description
for both reactions over a wide range of energies, i.e., from threshold to c.m. energies $W \leq 2$ GeV. This is important since the extractions of the resonances properties like mass, width,
and form factors would be reliable only from a consistent framework. Consistent extractions would  help to minimize model dependence such that comparison with LQCD
 results would be meaningful. Next is that the comparison of threshold results with  ChPT predictions wold offer  a glimpse of
the working of chiral symmetry. Lastly, since all dynamical models always assume a picture of bare quark core dressed by pion cloud, the success of the results could
help to understand the underlining dynamics of the structure of  nucleon resonances.

Among all the dynamical models available on the market, DMT model distinguishes itself in that it can describe all the data well from threshold, including the recent very
precise ones from MAMI and Jlab, up to  $W \leq$ 2 GeV, even though it includes only two channels. It was constructed over a period of about 20 years. It
started in 1985 when the dynamical approach was proposed in \cite{Yang85}, where it was shown that
with the use of separable $\pi N$ potential, resonable description of the S- and P-waves $\gamma\pi$ multipoles can be achieved within the dynamical approach.
When the experiments \cite{threshExp86} from Saclay and MAMI reported the violation of low-energy theorem for the $\pi^0$ threshold production, we demonstrated in \cite{Yang89}
that the violation could arise from the $\pi N$ final-state interaction (FSI), again with the use of separable $\pi N$ interaction. This prompted us, together with Harry Lee,
to undertake the task of constructing a realistic meson-exchange (MEX)  $\pi N$ potential. The use a MEX $\pi N$ model did bring the threshold values of $E_{0^+}(p\pi^0)$ quite close \cite{Lee91} to
the data even though  the energy dependence is not well reproduced. With the completion of Taipei-Argonne MEX $\pi N$ model \cite{Hung94,Hung01}, groups from Dubna and Mainz joined to imbed
the Taipei-Argonne MEX $\pi N$ model within the dynamical approach for $\pi N$ and $\gamma\pi$ reactions and extend the model to higher energies as well as   electroproduction
to become the DMT model \cite{KY99,Kamalov01a,Kamalov01b}.

In the followings, we will first present the basic formulation of the DMT model before summarizing the main results of the model. It will be seen that DMT model gives
excellent description of most of the available threshold data, including the recent ones from MAMI. Regarding the resonances properties, we will cover deformation of the
$\Delta(1232)$ and the extracted masses and widths of resonances up to 2 GeV and  compare them with PDG values.

\section{Formulation of the DMT Dynamical Model}
\subsection{$\pi N$ scattering}
\label{subsec2a}
For the $\pi N$ scattering, we start with the Bethe-Salpeter (BS) equation,
\begin{eqnarray}\label{BSeq}
T_{\pi N} = B_{\pi N} + B_{\pi N}G_0 T_{\pi N},
\end{eqnarray}

\noindent where $B_{\pi N}$ is the sum of all irreducible two-particle Feynman amplitudes
and $G_0$ the free relativistic $\pi N$ propagator.  The four-dimensional BS equation
can be reduced to a three-dimensional one by first recasting it  into following two equations,

\begin{eqnarray}\label{BSeqa}
 T_{\pi N} &=& \hat B_{\pi N} + \hat B_{\pi N}\hat G_0T_{\pi N},\\
\hat B_{\pi N} &= &B_{\pi N} + B_{\pi N}(G_0-\hat G_0)\hat B_{\pi N},
\end{eqnarray}
\noindent so that Eq. (2) would become three-dimensional with   an
appropriate choice of propagator $\hat G_0(k;P)$. It is important to choose $\hat
G_0$ such that two-body unitarity is maintained by reproducing the $\pi N$
elastic cut. There is a wide range of possible propagators that satisfy
this requirement. Since chiral symmetry plays an essential role in strong interaction, we employ the Cooper-Jennings
propagator \cite{CJ88} as it satisfies both  the soft pion theorems and unitarity. Furthermore,
we approximate $\hat B_{\pi N}$ by the tree diagrams of a chiral invariant Lagrangian
consisting of $\pi, N, \sigma, \rho,$ and $\Delta(1232)$ fields with pseudovector couplings, if interest is restricted from threshold  to
first resonance region.

For higher energies, we have to include the $\eta N$ channel as it is well-known that it couples strongly with
the two lowest $S_{11}$ resonances.  In addition, we introduce more resonances as dictated by the data. They are all bare and get
dressed by the meson cloud as in the case of $\Delta(1232)$. The effects of the $\pi\pi N$ channel are accounted for with introduction of a phenomenological width to
 the bare resonances. The details are given in \cite{Chen03,Chen07,Pascal07,Tiator10}.

The parameters in the model like coupling constants, cut-off in the form factors, and bare masses of the resonances are then varied
to obtain best fit to the phase shifts and inelasticities.
\subsection{$\gamma\pi$ reactions}
\label{subsec2b}
The dynamical approach to the EM production of pions starts with the following Lippman-Schwinger equation,
\begin{eqnarray}
t_{\gamma\pi}(E)=v_{\gamma\pi}+v_{\gamma\pi}g_0(E)\,t_{\pi
N}(E)\,,\label{eq:tmatrix}
\end{eqnarray}
where $v_{\gamma\pi}$ is the $\gamma\pi$ transition potential,
$g_0$ and $t_{\pi N}$ are the $\pi N$ free propagator and
$t-$matrix, respectively, and $E$ is the total energy in the c.m.
frame.   $t_{\pi N}$
is related to   $T$ defined in Eq. (\ref{BSeq}) by some kinematical factor.
$v_{\gamma\pi}$  are derived from an  effective Lagrangian obtained from gauging
the chiral invariant Lagrangian used for the $\pi N$ scattering. It
contains Born terms as well as $\rho-$ and $\omega-$exchange in
the $t-$channel \cite{Olsson75}. For  electroproduction,
 gauge invariance is restored by the substitution,
$J_{\mu} \rightarrow J_{\mu}  - k_{\mu}\frac{k\cdot J}{k^2}\,,$
 where $J_{\mu}$ is the electromagnetic current corresponding to
the background  contribution of $v^B_{\gamma\pi}$.

For the physical multipoles in channel $\alpha=\{l,j\}$, Eq.
(\ref{eq:tmatrix}) gives~\cite{Yang85}
\begin{eqnarray}
t_{\gamma\pi}^{\alpha}(q_E,k)=\exp{(i\delta^{\alpha})}\,\cos{\delta^{\alpha}}
\left[ v_{\gamma\pi}^{\alpha}(q_E,k) + P\int_0^{~} dq'
\frac{R_{\pi N}^{\alpha}(q_E,q')\,v_{\gamma\pi}^{\alpha}(q',k)}{E(q_E)-E(q')}\right]\,,
\label{eq:Tback}
\end{eqnarray}
where $\delta^{\alpha}$ and $R^{\alpha}$ are the $\pi N$ phase
shift and  reaction matrix, in channel $\alpha$, respectively,
$q_E$ is the pion on-shell momentum and $k=\mid {\bf k}\mid$ the
photon momentum. Note that the second term on the r.h.s. of the principal integral term in
Eq. (\ref{eq:Tback}), depends on the off-energy-shell behaviors of both $R_{\pi N}^{\alpha}$ and $v_{\gamma\pi}^{\alpha}$.
To maintain gauge invariance for the off-energy-shell matrix elements of $v_{\gamma\pi}^{\alpha}$ is a nontrivial task and
different groups have followed different recipes. The prescription we used are expounded in \cite{Pascal07}.

\section{Results and Discussions}
\label{sec3}
In this section we summarize some of the featured results of DMT model. This includes threshold $\pi^0$ production, $\Delta(1232)$ deformation, and
the extracted properties of the higher resonances.
\subsection{Threshold $\pi^0$ production}
\label{subsec3a}
The discrepancy between  experiments \cite{threshExp86} and the  prediction of low-energy theorem based on current algebra and PCAC,
for the $\pi^0$ threshold production came as a big surprise. First it was suggested \cite{Yang89} that the discrepancy could arise from the $\pi N$
final-state interaction (FSI). Soon it was recognized the EM threshold $\pi^0$ production provides an excellent arena to study spontaneous
as well as explicit chiral symmetry breaking with latter arising from the non-vanishing
small masses of $u$ and $d$ quarks. This has spurred  intensive theoretical and experimental efforts which persist until today.

On the theoretical side, two different approaches have been undertaken, namely, dynamical model and  chiral perturbation
theory (ChPT)\cite{7thChiral}.
In the followings, results from DMT model, the most successful dynamical model for the threshold pion production, are compared with predictions of
relativistic chiral perturbation theory (RChPT)\cite{Hilt13,Scherer15}, heavy baryon chiral perturbation theory (HBChPT)\cite{Bernard96}, and the latest measurements from MAMI.

%%%%%%%%%%%%%%%%%%%%%%%%%%%%%%%%%%%%%%%%%%%%%%%%%%
%%%%%%%%%%%%%%%%%%%%%%%%%%%%%%%%%%%%%%%%%%%%%%%%%%
\begin{figure}[tbh]
\begin{minipage}{0.56\linewidth}
We  show in Fig. 1, the real  and imaginary   parts of the $E_{0+}(p\pi^0)$ near threshold. The blue dashed curve denotes
 the result obtained without FSI in DMT. The black solid
and red dot-dot-dashed  lines correspond to the predictions of the full DMT model and the HBChPT, respectively. The data are from \cite{threshExp86}. It is seen that
both DMT model and HBChPT can describe the data well. The green dot-dashed
and purple dotted curves are the results obtained with one-loop and two-loop corrections included within DMT model, namely, $t_{\gamma\pi}(E)$ is approximated
with $v_{\gamma\pi}+v_{\gamma\pi}g_0(E)\,v_{\pi N}(E)$ and $v_{\gamma\pi}+v_{\gamma\pi}g_0(E)\,v_{\pi N}(E)+v_{\gamma\pi}g_0(E)\,v_{\pi
N}(E)g_0(E)\,v_{\pi N}(E)$, respectively. One sees that
\end{minipage}
\hfill
\begin{minipage}{0.40\linewidth}
\center{\includegraphics[width=0.87\textwidth]{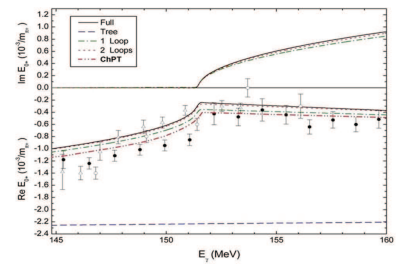}}
\center{\caption{Comparison of  predictions of DMT model,
HBChPT with data for $E_{0^+}(p\pi^0)$ near threshold. See text for notations.}}
\end{minipage}
\label{fig1}
\end{figure}
\noindent the two-loop corrections are small and ChPT
 calculations is justified to stop within
the one-loop scheme.
%%%%%%%%%%%%%%%%%%%%%%%%%%%%%%%%

Fig. 2 shows the coincidence cross sections $\sigma_0, \sigma_{TT},$ and $\sigma_{LT}$ in $\mu b/sr$ and beam asymmetry
$A_{LT}$ in $\%$ measured at MAMI \cite{Weis08} in $\pi^0$ electroproduction  at constant $Q^2 = 0.05$ GeV$^2, \Theta_\pi = 90^\circ,
\Phi_\pi = 90^\circ,$ and $\epsilon = 0.93$ as a function of
$\Delta W$ above threshold. The red solid lines show  RChPT calculations at $O(q^4)$ and the black dotted
 lines are the heavy-baryon ChPT calculations of \cite{Bernard96}. The green dashed  curves are
obtained from  DMT model.  It is seen that the data are in disagreement with predictions of HBChPT,
while in good agreement with DMT model and RChPT predictions with DMT doing somewhat a better job.
In Fig. 3, the polarised differential cross sections $\sigma_T$ of threshold $\pi^0$ photoproduction with transverse polarized protons measured at MAMI
for photon energies at 168.5 and 183.7 MeV, respectively,
are depicted \cite{Schumann15}. Data points represent
Crystal Ball/TAPS results with statistical uncertainties only. Solid lines are predictions of the DMT model, while dashed and
dashed-dotted lines show three-parameter Legendre fits to the experimental data and the cross-check analysis, respectively. Again DMT model describes
these data very well.

The success of DMT model in describing EM $\pi^0$ production near threshold can be understood as follows. In Eq. (\ref{eq:tmatrix}), it is
seen that the $t_{\gamma\pi}$ depends only on $v_{\gamma\pi}$ and $t_{\pi N}$. In the threshold region, only background mechanisms contribute,
namely, $v_{\gamma\pi}$ and $t_{\pi N}$ would  be given by $v^B_{\gamma\pi}$ and $t^B_{\pi N}$, respectively, where superscript $B$ refers to background mechanism.
 Since $v^B_{\gamma\pi}$ and $v^B_{\pi N}$, which drives $t^B_{\pi N}$, are both derived from tree diagrams of chiral invariant effective Lagrangian, and we use Cooper-Jennings' propagator which satisfies soft-pion theorem to generate $t^B_{\pi N}$, it is not surprising that the resulting
$t^B_{\gamma\pi}$ would preserve many of the consequences of chiral symmetry.

%%%%%%%%%%%%%%%%%%%%%%%%%%%%%%%%%%%%%%%%%
\begin{figure}[tbh]
\begin{minipage}{0.48\linewidth}
\center{\includegraphics[width=0.95\textwidth]{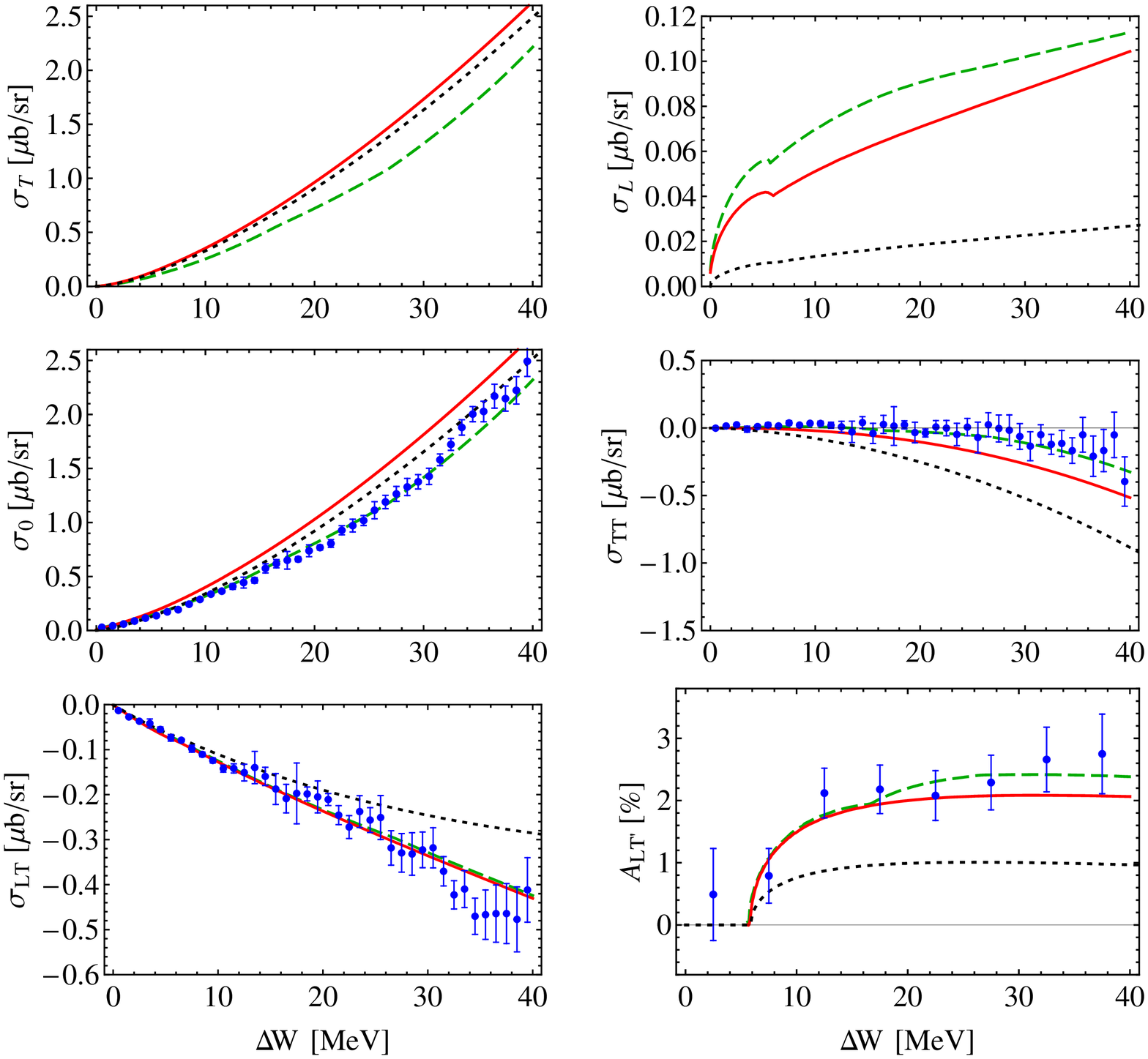}}
\center{\caption{$\sigma_0, \sigma_{TT},$ and $\sigma_{LT}$  and
$A_{LT}$   at  $Q^2 = 0.05$ GeV$^2$, $\Theta_\pi = 90^\circ, \Phi_\pi = 90^\circ$, and $\epsilon = 0.93$ vs.
$\Delta W$. See text for notations.}}
\end{minipage}\label{fig2}
\hfill
\begin{minipage}{0.48\linewidth}
\center{\includegraphics[width=0.55\textwidth]{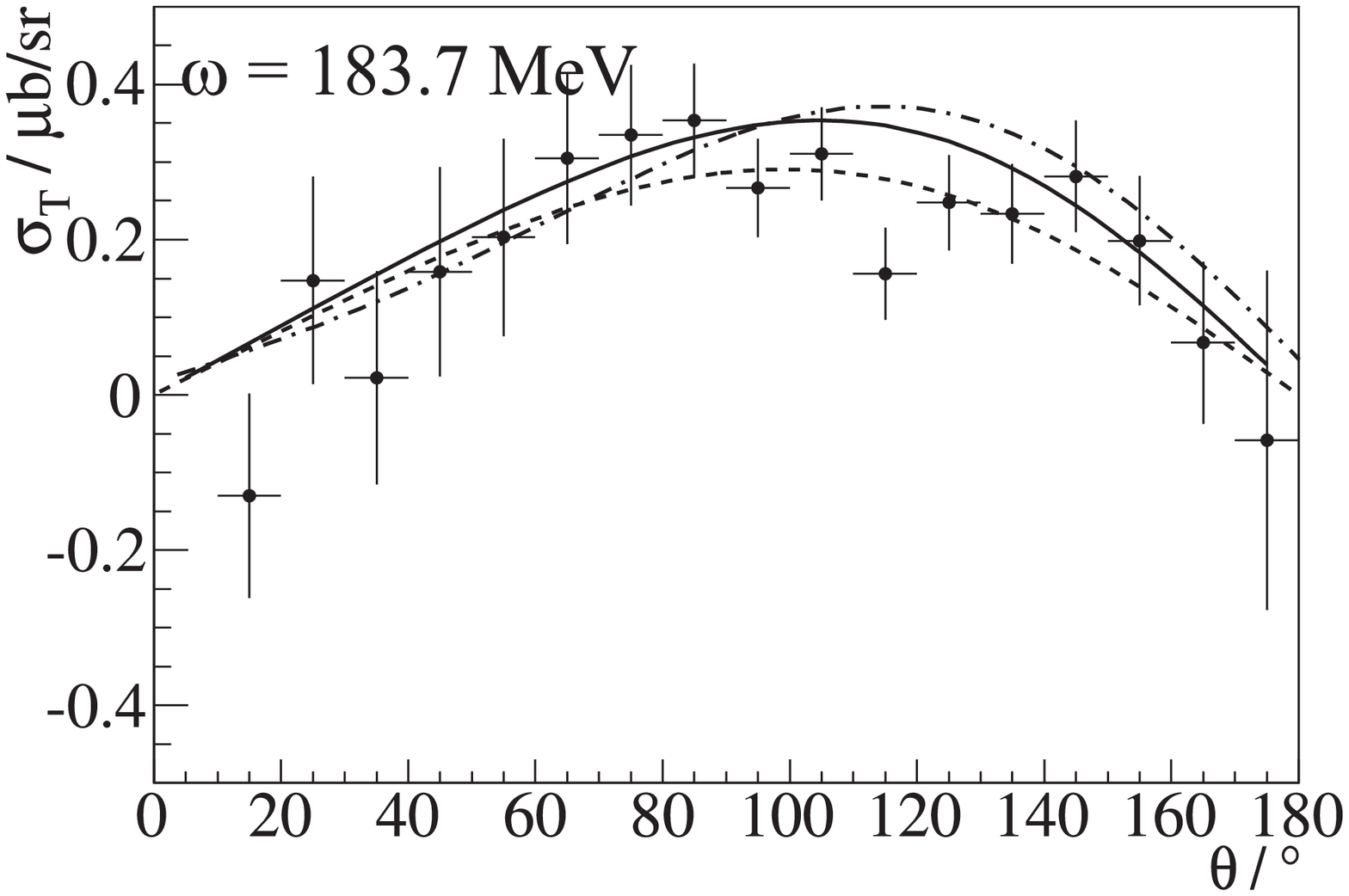}}
\center{\includegraphics[width=0.55\textwidth]{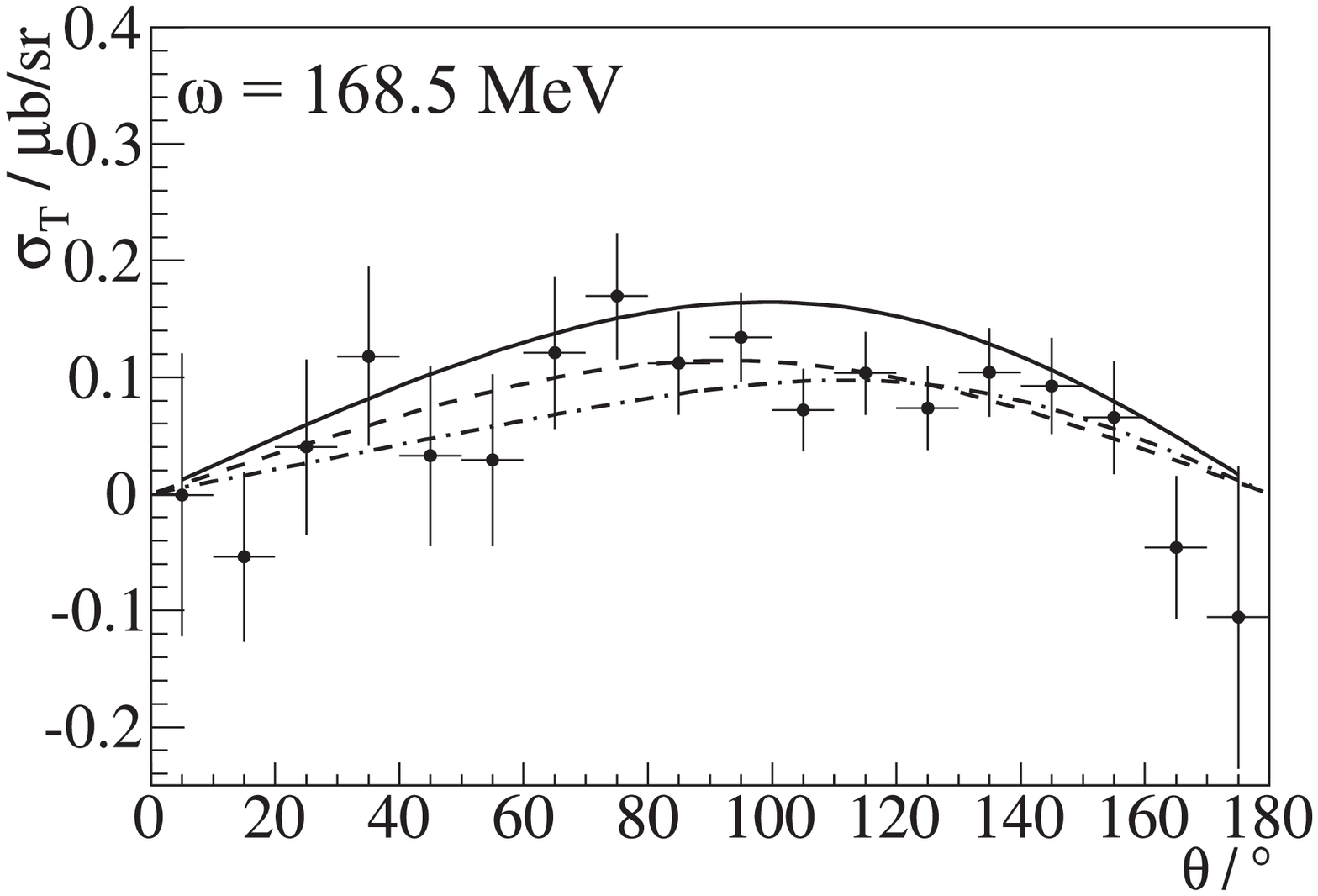}}
\center{\caption{
$\sigma_T$. Data points represent Crystal Ball /TAPS results with statistical uncertainties only. See text for notations.
}}
\end{minipage}
\label{fig3}
\end{figure}

\subsection{(3,3) Multipoles $M_{1^+}, E_{1^+}$ and the Deformation of $\Delta(1232)$}
\label{subsec3b}
In the begining, the dynamical approach was proposed for the first resonance region \cite{Tanabe85,Yang85} in order
to unitarize the $\gamma\pi$ multipoles to satisfy the Fermi-Watson theorem dynamically. In the (3,3) channel where $\Delta$ excitation plays an important role,
the transition potential $v_{\gamma\pi}$ consists of two terms
\begin{eqnarray}
v_{\gamma\pi}(E)=v_{\gamma\pi}^B + v_{\gamma\pi}^{\Delta}(E)\,,
\label{eq:tranpot}
\end{eqnarray}
where $v_{\gamma\pi}^B$ is the background transition potential
which, as discussed earlier, includes Born terms and vector mesons exchange
contributions.  The second term of Eq. (\ref{eq:tranpot})
corresponds to the contribution of bare $\Delta$,
namely,  $\gamma^* N \rightarrow \Delta \rightarrow \pi N$. The vertex $\gamma^* N \rightarrow \Delta$
would introduce three more parameters corresponding to $M1, E2,$   and  $C2$
 excitations.

We proceed by decomposing Eq. (\ref{eq:tmatrix}) in the following way,
\begin{eqnarray}
t_{\gamma\pi}= t_{\gamma\pi}^B + t_{\gamma\pi}^{\Delta}\,,
\label{eq:decomp}
\end{eqnarray}
where
\begin{eqnarray}
t_{\gamma\pi}^B(E)=v_{\gamma\pi}^B+v_{\gamma\pi}^B\,g_0(E)\,t_{\pi
N}(E),\label{t-gampi-b}\\
t_{\gamma\pi}^\Delta(E)=v_{\gamma\pi}^\Delta+v_{\gamma\pi}^\Delta\,
g_0(E)\,t_{\pi N}(E).\label{t-gampi-delta}
\end{eqnarray}
The advantage of such a decomposition (\ref{eq:decomp}) is that both $t_{\gamma\pi}^B$
and $t_{\gamma\pi}^\Delta$ would satisfy the Fermi-Watson theorem as can be inferred
from Eq. (\ref{eq:Tback}). $t_{\gamma\pi}^\Delta$ would contain
all the processes which start with the electromagnetic excitation
of the bare $\Delta$.  It provides   a prescription to extract
information concerning bare $\Delta$ excitation.

 With $t_{\pi N}$ obtained as prescribed in Sec.~\ref{subsec2a}, $t_{\gamma\pi}^B$ can be calculated
straightforwardly. For real

%%%%%%%%%%%%%%%%%%%%%%%%%%%%%%%%%%%%%%%%%%%%%%%%%%
\begin{figure}[tbh]
\begin{minipage}{0.50\linewidth}
 photon, $t_{\gamma\pi}^\Delta$ depends on the $M1$ and $E2$ excitation strength
of $\gamma N \rightarrow \Delta$. By combining the
contributions of $t_{\gamma\pi}^B$ and $t_{\gamma\pi}^\Delta$ with the excitation strength as free
parameters, results of our best fit to the  real and imaginary  parts of the $M_{1+}^{(3/2)}$ and $E_{1+}^{(3/2)}$ multipoles
obtained in
the  analyses of Mainz~\cite{HDT} and VPI group~\cite{VPI97}
are shown in Fig. 4.    The
dashed (dotted) curves are the results obtained within the DMT
model for $t_{\gamma\pi}^B$ including (excluding) the principal
value integral contribution in Eq. \ref{eq:Tback}. The solid curves are
the full DMT results including also the bare $\Delta$ excitation.
For the $E_{1+}$ multipole, the dashed and solid curves
practically coincide due to the small value of the bare $\Delta$ contribution. The open and solid circles are the results from the Mainz
 and   VPI analyses, respectively. In other words, The pruple and yellow regions represented
the the contributions from bare $\Delta$ excitation and the off-energy-shell $\pi N$ rescatterings associated with pion cloud.

\end{minipage}
\hfill
\begin{minipage}{0.46\linewidth}
\center{\includegraphics[width=0.87\textwidth]{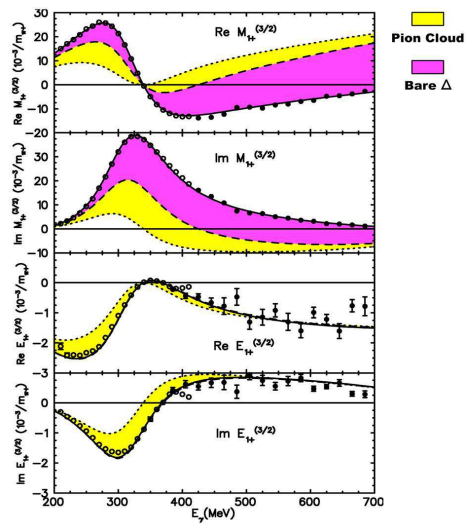}}
\center{\caption{Our results for $M_{1+}^{(3/2)}$ and $E_{1+}^{(3/2)}$ multipoles.  See text for notations. Figure from  \cite{KY99}.}}
\end{minipage}
\label{fig1}
\end{figure}
%%%%%%%%%%%%%%%%%%%%%%%%%%%%%%%%

  At low four-momentum transfer squared $Q^2$,  the interest in the EM excitation of $\Delta$ lies in the observation of
a $D-$state in  $\Delta$ \cite{Beck97,Hsiao98,Tiator03}. It would indicate that   $\Delta$ is deformed and
the photon can excite a nucleon via  $E2$ and
$C2$  transitions. In a symmetric SU(6) quark model,
the electromagnetic excitation of  $\Delta$ could proceed only
via $M1$ transition. In pion electroproduction, $E2$ and $C2$
excitations would give rise to nonvanishing $E_{1+}^{(3/2)}$ and
$S_{1+}^{(3/2)}$ multipole amplitudes. Currently, summary
of experiments give, near $Q^2 =0$ and at $W=1232$ MeV, $R_{EM} = E_{1+}^{(3/2)}/M_{1+}^{(3/2)} = -(2.5\pm0.5)\%$
\cite{PDG06}, whereas we obtained a value of $0.24\%$, a clear indication of $\Delta$ deformation.

Our results shown in Fig. 4 offer an interesting dynamical picture for the $\Delta(1232)$ deformation. Namely, bare
$\Delta$ contribution to $E_{1^+}$, as  would be denoted in red,   nearly vanishes. It implies that the bare $\Delta$ is mostly spherical
as would be in a symmetric SU(6) quark model. The $\Delta$ deformation arises from the dressing of pion cloud, as represented in yellow
area for $E_{1^+}$. It corresponds to the   principal
value integral contribution in Eq. (\ref{eq:Tback}) and describes the off-energy-shell $\pi N$ rescattering effects.

\subsection{Extracted Properties of Higher Resonances}
\label{subsec3c}
The meson-exchange $\pi N$ model described in Sec.~\ref{subsec2a}. has been extended to energies $W\leq 2$ GeV \cite{Chen07} and used to extract
  properties of higher resonances.   We illustrate our scheme for $S_{11}$ channel in the followings,
where we   include the $\eta N$
channel and enlarge the Hilbert space to accommodate as many resonances as
would be required by the data. We assume that each contributing bare resonance
$R$   acquires a width by  coupling to  $\pi N$ and $\eta N$ channels.
Details can be found in \cite{Chen07}.

The full $t$-matrix can  be written as a system of coupled equations,
\begin{equation}
t_{ij}(E)= v_{ij}(E)+\sum_k  v_{ik}(E)\,g_k(E)\, t_{kj}(E)\,, \label{eq:3.2.2}
\end{equation}
with $i$ and $j$ denoting the $\pi$ and $\eta$ channels and $E\equiv W$, the total c.m. energy . The potential $ v_{ij}$ is, as in Eq. (\ref{eq:tranpot}), a sum of background $(v^B_{ij})$
and bare resonance $(v^R_{ij})$ terms,
$v_{ij}(E)=  v^B_{ij}(E)+ v^R_{ij}(E),$
where $v^B_{\pi\pi}$ for the $\pi N$ elastic channel is taken
as obtained in Sec.~\ref{subsec2a}.  In  channels involving $\eta$, the
potential $v^B_{i\eta}$ is assumed to vanish
because of the small $\eta NN$ coupling.

The bare resonance contribution arises from  excitation and decay of the
resonance $R$.
 The matrix elements of the potential $v^R_{ij}(E)$
can be symbolically expressed by
\begin{equation}
v^R_{ij}(q,q';E)=\frac{f_i(\tilde {\Lambda}_i,q;E)\,g_i^{(0,R)}\,g_j^{(0,R)}\,
f_j(\tilde{\Lambda}_j,q';E)}{E-M_R^{(0)}+ \frac{i}{2}\Gamma_R^{2\pi}(E)} \,,
\label{eq:3.2.5}
\end{equation}
where $M_R^{(0)}$ denotes the mass of bare resonance $R$; $q$ and $q'$ are the
pion (or eta) momenta in the initial and final states, and $g_{i/j}^{(0,R)}$
denotes the resonance vertex couplings. $\tilde\Lambda_i$ stands for a triple
of cut-offs, ($\Lambda_N,\Lambda_R,\Lambda_\pi$), defined in form factor
$F_{\alpha}(p^2_{\alpha})= \left[ n\Lambda_{\alpha}^4/\left(n\Lambda_{\alpha}^4
+ (m_{\alpha}^2 - p_{\alpha}^2)^2\right)\right]^{n}$,
with $p_{\alpha}$ the four-momentum and $m_{\alpha}$ the mass of particle
$\alpha$.
In  Eq.~(\ref{eq:3.2.5}), we have included a
phenomenological term $\Gamma_R^{2\pi}(E)$ in the resonance propagator to
account for the $\pi\pi N$ decay channel. Therefore, our bare resonance
propagator already contains a phenomenological ``dressing'' effect due to the
coupling to   $\pi\pi N$ channel. We use the parameterization of
\cite{Lvov97,Drechsel99} which has the
correct threshold behavior and contains another cut-off.  With this prescription we assume that any
further non-resonant coupling to the $\pi\pi N$ channel can be neglected. All together,
each resonance is generally described by 6 free
parameters.

 The generalization of the coupled-channel model to the case
of $N$ resonances with the same quantum numbers is then given by
$v^R_{ij}(q,q';E)=\sum_{n=1}^{N} v^{R_n}_{ij}(q,q';E)$.
The solutions of the coupled-channel equations of Eq.~(\ref{eq:3.2.2}), with
potentials given above were fitted to the $\pi N$ phase shifts and
inelasticity  in all channels up to the $F$-waves and
for $W\leq 2$ GeV.  We obtain  excellent
description for both the real and the imaginary parts of the
pion-nucleon scattering amplitudes in all cases except for the
$D_{35}$ and $F_{17}$ channels.   However, the fit to the data requires four additional resonances with
very large widths, $S_{11}(1878), D_{13}(2152), P_{13}(2204)$, and
$P_{31}(2100)$, which are not listed by the PDG~\cite{PDG06}.

The physical mass $M_R$, total width $\Gamma_R$, single-pion
branching ratio $\beta_R^{1\pi}$, and background phase ${\phi_R}$
defined for each overlapping nucleon resonance $R$ have been
determined as explained in \cite{Chen07}.
$T$-matrix poles have been calculated by three different
techniques: analytic continuation into the complex energy plane, speed-plot,
and regularization method \cite{Tiator10}.
Only the results for bare ($M_R^{(0)}$) and
physical ($M_R$) resonance masses and total widths $\Gamma_R$ are
listed in the following Tables I and II and compared with listings of PDG, for
the isospin-$\frac{3}{2}$ and isospin-$\frac{1}{2}$ resonances,
respectively.  More results can be found in \cite{Chen07}. One sees a
qualitative agreement in general but considerable discrepancies in some cases,
in particular for the widths  of some higher resonances. Further
investigations will be necessary to understand these differences in detail.

%%%%%%%%%%%%%%%%%%%%%%%%%%%%%%%%%%%%%%%%%%%%%%%%%%%%%%%%%%%%%%%%%%%%%%%%%%%%%%%%%%%%%%%%%

\begin{table}[h]
\begin{minipage}{0.48\linewidth}
 \centering \setlength{\textwidth}{50mm} \caption{Bare ($M_R^{(0)}$) and
physical ($M_R$) resonance masses and total widths $\Gamma_R$, all in
units of MeV,  for $I=3/2$ resonances. Upper lines: our results, lower
lines: PDG values  \cite{PDG06}.}
\begin{tabular}{|l|ccc|}
\hline
 $N^*$ & $M_R^{(0)}$ & $M_R$ & $\Gamma_R$  \\
\hline
$P_{33}(1232)$  & 1425 & 1233 & 132  \\
 $****$&  &$1232\pm 1$& $118\pm 2$   \\
 \hline
$P_{33}(1600)$  & 1575 & 1562 & 216    \\
 $***$ &  &$1600\pm 100$& $350\pm 100$   \\
 \hline
$S_{31}(1620)$  & 1654 & 1616 & 160   \\
 $****$ &  &$1630\pm 30$& $142\pm 18$   \\
 \hline
$D_{33}(1700)$  & 1690 & 1650 & 260   \\
 $****$ &  &$1710\pm 40$& $300\pm 100$  \\
\hline
$P_{31}(1750)$  & 1765 & 1746 & 554   \\
 $*$  &  &$1744\pm 36$& $300\pm 120$   \\
 \hline
$S_{31}(1900)$  & 1796 & 1770 & 430   \\
 $**$ &  &$1900\pm 50$& $190\pm 50$   \\
 \hline
$F_{35}(1905)$  & 1891 & 1854 & 534   \\
 $****$  &  &$1890\pm 25$& $335\pm 65$   \\
 \hline
$P_{31}(1910)$  & 1953 & 1937 & 226   \\
 $****$   &  &$1895\pm 25$& $230\pm 40$   \\
 \hline
$P_{33}(1920)$  & 1856 & 1827 & 834    \\
 $***$   &  &$1935\pm 35$& $220\pm 70$   \\
 \hline
$D_{35}(1930)$  & 2100 & 2068 & 426  \\
 $***$  &  &$1960\pm 60$& $360\pm 140$   \\
 \hline
$D_{33}(1940)$  & 2100 & 2092 & 310    \\
 $*$  &  &$2057\pm 110$& $460\pm 320$   \\
 \hline
$F_{37}(1950)$  & 1974 & 1916 & 338   \\
 $****$   &  &$1932\pm 17$& $285\pm 50$   \\
 \hline
$F_{35}(2000) $  & 2277 & 2260 & 356   \\
 $**$   &  &$2200\pm 125$& $400\pm 125$   \\
 \hline
$P_{31}(xxx)$  &  2160 & 2100 & 492   \\
 \hline
$S_{31}(2150)$  & 2118 & 1942 & 416   \\
 $*$  &  &$2150\pm 100$& $200\pm 100$  \\
\hline

\end{tabular}
\end{minipage}
\hfill
\begin{minipage}{0.48\linewidth}
\centering \setlength{\textwidth}{50mm} \caption{$I=1/2$ resonances. Notations same
as in Table I.}
\begin{tabular}{|l|ccc|}
\hline
 $N^*$ & $M_R^{(0)}$ & $M_R$ & $\Gamma$  \\
\hline
$P_{11}(1440)$  & 1612 & 1418 & 436    \\
 $****$&  &$1445\pm 25$& $325\pm 125$   \\
 \hline
$D_{13}(1520)$  & 1590 & 1520 &94   \\
 $****$ &  &$1520\pm 5$& $115\pm 15$   \\
 \hline
$S_{11}(1535)$  & 1559 & 1520 & 130   \\
 $****$ &  &$1535\pm 10$& $150\pm 25$   \\
 \hline
$S_{11}(1650)$  & 1727 & 1678 & 200    \\
 $****$ &  &$1655\pm 10$& $165\pm 20$   \\
\hline
$D_{15}(1675)$  & 1710 & 1670 & 154   \\
 $****$  &  &$1675\pm 5$& $147\pm 17$   \\
 \hline
$F_{15}(1680)$  & 1748 & 1687 & 156   \\
 $****$ &  &$1685\pm 5$& $130\pm 10$   \\
 \hline
$D_{13}(1700)$  & 1753 & 1747 & 156   \\
 $***$  &  &$1700\pm 50$& $100\pm 50$   \\
 \hline
$P_{11}(1710)$  & 1798 & 1803 & 508    \\
 $***$   &  &$1710\pm 30$& $180\pm 100$   \\
 \hline
$P_{13}(1720)$  & 1725 & 1711 & 278    \\
 $****$   &  &$1725\pm 25$& $225\pm 75$   \\
 \hline
$P_{13}(1900)$  & 1922 & 1861 & 1000  \\
 $**$  &  &$1879\pm 17$& $498\pm 78$   \\
 \hline
$F_{15}(2000)$  & 1928 & 1926 & 58    \\
 $**$  &  &$1903\pm 87$& $490\pm 310$  \\
 \hline
$D_{13}(2080)$  & 1972 & 1946 & 494    \\
 $**$   &  &$1804\pm 55$& $450\pm 185$   \\
 \hline
$S_{11}(xxx) $  & 1803 & 1878 & 508    \\
 \hline
$S_{11}(2090)$  & 2090 & 2124 & 388   \\
 $*$   &  &$2180\pm 80$& $350\pm 100$   \\
 \hline
$P_{11}(2100)$  & 2196 & 2247 & 1020   \\
 $*$  &  &$2125\pm 75$& $260\pm 100$   \\
 \hline
$D_{13}(xxx) $  & 2162 & 2152 & 292   \\
\hline
$P_{13}(xxx )$  & 2220 & 2204 & 406   \\
\hline
$D_{15}(2200)$  & 2300 & 2286 & 532    \\
 $**$ &  &$2180\pm 80$& $400\pm 100$    \\
 \hline
\end{tabular}
\end{minipage}
\end{table}

%%%%%%%%%%%%%%%%%%%%%%%%%%%%%%%%%%%%%%%%%%%%%%%%%%%%%%%%%%%%%%%%

%
\section{Summary}
Some of the featured results of the Dubna-Mainz-Taipei (DMT) dynamical model  for $\pi N$ scattering and $\pi^0$ electromagnetic (EM) production are summarized.
These include results for  threshold production, deformation of $\Delta(1232)$, and the extracted   properties,  including  masses and widths,
of higher resonances below 2 GeV. The excellent agreement of DMT model's predictions with threshold $\pi^0$ production data, including the recent precision measurements
from MAMI  establishes results of the DMT model as a benchmark \cite{DMT} for experimentalists and theorists in dealing with threshold pion production.
In the first resonance region with low momentum transfer, DMT model provides an excellent description of the existing data and offers a dynamical picture for
the $\Delta(1232)$  deformation. Namely, the bare $\Delta$ is almost spherical and the deformation of   physical $\Delta$ arises from the dressing of the
core by pion cloud.

Going beyond the first resonance region but below $W\leq$ 2 GeV, the model was extended
 by including the $\eta N$ channel and all
the $\pi N$ resonances with masses $\leq 2$ GeV, up to the $F$
waves. The effects of the $\pi\pi N$ channels are taken into
account by introducing an effective width in the resonance
propagators. The extended model gives an excellent fit to both
$\pi N$ phase shifts and inelasticity parameters in all channels
up to the $F$ waves and for energies below 2 GeV. However, the fit to the data requires four additional resonances with
very large widths, $S_{11}(1878), D_{13}(2152), P_{13}(2204)$, and
$P_{31}(2100)$, which are not listed by PDG~\cite{PDG06}.
In addition, the
predicted values for the resonance masses and widths are compared to the listing
of the PDG. In general, there is
qualitative agreement   but considerable discrepancies exist in some cases,
in particular for the widths  of some higher resonances. Further
study will be necessary to understand these differences in detail.
\\

{\noindent \bf Acknowledgement} \\
\\
Results summarized here have been obtained in collaborations with G.Y. Chen, D. Dreschel, C.T. Hung, S.S. Kamalov, C.C. Lee, T.-S.H. Lee, and L. Tiator. I dedicate
this article to the memory of Mr. Chau-Chen Lee, the first author of Ref. \cite{Lee91}, who recently passed away abruptly. I am indebted to L. Tiator for providing me
with some of the figures shown.

\end{document}